\documentclass[journal]{IEEEtran}
\usepackage{cite}
\usepackage{enumitem}
\usepackage{amsmath,amssymb,amsfonts,amsthm}
\usepackage{algorithmic}
\usepackage{graphicx}
\usepackage{textcomp}
\usepackage{xcolor}
\usepackage{extdash}
\ifCLASSOPTIONcompsoc \usepackage[caption=false,font=normalsize,labelfont=sf,textfont=sf]{subfig}
\else
\usepackage[caption=false,font=footnotesize]{subfig}
\fi

\usepackage{mathtools}
\usepackage{verbatim}
\def\BibTeX{{\rm B\kern-.05em{\sc i\kern-.025em b}\kern-.08em
		T\kern-.1667em\lower.7ex\hbox{E}\kern-.125emX}}

\usepackage[acronym,nomain, shortcuts, nopostdot]{glossaries}\glstoctrue
\usepackage{microtype}

\newcommand{\mrm}{\mathrm}

\newtheorem{remark}{Remark}
\newtheorem{assumption}{Assumption}

\newtheorem{theorem}{Theorem}
\newcommand{\nlm}[1]{{\color{red}{}#1}}

\newcommand{\Nsw}{\mathcal{N}^\mrm{sg}}

\newcommand{\nsw}{n}
\newcommand{\omegacoi}{\omega^\mrm{coi}}

\newcommand{\omegasw}{\omega}
\newcommand{\Domegasw}{\Delta\omega_i}
\newcommand{\domegasw}{\dot{\omega}}
\newcommand{\Hsw}{H}
\newcommand{\Hca}{H^\mrm{ca}}

\newcommand{\Htot}{H^\mrm{tot}}
\newcommand{\hHtot}{\hat{H}^\mrm{tot}}
\newcommand{\Ptswi}{p_{\mrm{t},i}}
\newcommand{\Peswi}{p_{\mrm{e},i}}
\newcommand{\Pmswi}{p_{\mrm{m},i}}
\newcommand{\Petotj}{p_{\mrm{e},j}^\mrm{ca}}
\newcommand{\Pmtotj}{p_{\mrm{m},j}^\mrm{ca}}
\newcommand{\Petot}{p_{\mrm{e}}^\mrm{tot}}
\newcommand{\Pmtot}{p_{\mrm{m}}^\mrm{tot}}
\newcommand{\deltasw}{\delta}
\newcommand{\ddeltasw}{\dot{\delta}}

\newcommand{\Peloadj}{p_{\mrm{load},j}^\mrm{ca}}
\newcommand{\Pelossj}{p_{\mrm{loss},j}^\mrm{ca}}
\newcommand{\Peresj}{p_{\mrm{res},j}^\mrm{ca}}
\newcommand{\Peiaj}{p_{\mrm{in},j}^\mrm{ca}}

\newcommand{\aca}{a^\mrm{ca}}
\newcommand{\atot}{a^\mrm{tot}}
\newcommand{\hatot}{\hat{a}^\mrm{tot}}
\newcommand{\dxcoi}{\dot{\omega}^\mrm{coi}}

\newcommand{\xav}{{\omega}^\mrm{av}}
\newcommand{\dxav}{\dot{\omega}^\mrm{av}}

\newcommand{\dxsw}[1][]{%
	\ifthenelse{\equal{#1}{} }
	{\ensuremath{\dot{x}_i}}
	{\ensuremath{\dot{x}_{#1,i}}}
}
\newcommand{\xsw}[1][]{%
	\ifthenelse{\equal{#1}{} }
	{\ensuremath{{x}_i}}
	{\ensuremath{{x}_{#1,i}}}
}
\newcommand{\Fav}{\mathcal{F}}
\newcommand{\lambdaav}[1]{\lambda_{#1}}
\newcommand{\lambdaavi}{\lambda_{i}}
\newcommand{\thetah}{\theta}
\newcommand{\yh}{y}
\newcommand{\nuh}{\nu}
\newcommand{\Ch}{c}
\newcommand{\hthetah}{\hat{\theta}}
\newcommand{\dhthetah}{\dot{\hat{\theta}}}

\newcommand{\Gammah}{\Gamma}
\newcommand{\alphah}{\alpha}

\newcommand{\bGammahopt}{\bar{\Gamma}^*}
\newcommand{\alphahopt}{\alpha^*}

\newcommand{\Ghcal}{\mathcal{G}}
\newcommand{\Vhcal}{\mathcal{V}}
\newcommand{\Ehcal}{\mathcal{E}}
\newcommand{\Nhcal}{\mathcal{N}}
\newcommand{\nhe}{n_\mrm{e}}
\newcommand{\Dh}{D}
\newcommand{\Lh}{L}

\newcommand{\Timeh}{T}
\newcommand{\cPEuh}{\bar{\iota}_1}
\newcommand{\cPElh}{\underline{\iota}_1}
\newcommand{\rh}{r}
\newcommand{\lambdah}[1]{\lambda_{#1}}

\newcommand{\lambdahk}{\lambda_{k}}

\newcommand{\lambdahunder}{\underline{\lambda}}

\newcommand{\xh}{x}
\newcommand{\hxh}{\hat{x}}
\newcommand{\txh}{\tilde{x}}
\newcommand{\dtxh}{\dot{\tilde{x}}}
\newcommand{\onensw}{1_{n_\mrm{a}}}
\newcommand{\bGammah}{\bar{\Gamma}}

\newcommand{\bCh}{\bar{C}}
\newcommand{\bLa}{\bar{\Lambda}}
\newcommand{\ytilh}{\tilde{y}}
\newcommand{\yglobalh}{y}
\newcommand{\yhath}{\hat{y}}

\newcommand{\Nsg}{N}
\newcommand{\Rsg}{R_\mrm{s}}

\newacronym{wams}{WAMS}{wide area measurement system}
\newacronym{db}{dB}{decibels}
\newacronym{smape}{sMAPE}{symmetric mean absolute percentage error}
\newacronym{snr}{SNR}{signal to noise ratio}
\newacronym{ode}{ODE}{ordinary differential equation}
\newacronym{bibo}{BIBO}{bounded-input bounded-output}
\newacronym{drem}{DREM}{Dynamic Regressor Extension and Mixing}
\newacronym{dsa}{DSA}{dynamic security assesment}
\newacronym{lti}{LTI}{linear time-invariant}
\newacronym{der}{DER}{distributed energy resource}
\newacronym{dse}{DSE}{power system dynamic state estimation}
\newacronym{kf}{KF}{Kalman filter}
\newacronym{kfo}{KFo}{Koalman filter}
\newacronym{pmu}{PMU}{Phasor Measurement Unit}
\newacronym{tso}{TSO}{transmission system operator}
\newacronym{avr}{AVR}{automatic voltage regulator}
\newacronym{pss}{PSS}{power system stabilizer}
\newacronym{sg}{SG}{synchrounous generator}
\newacronym{linre}{LRE}{linear regression equation}
\newacronym{ltv}{LTV}{linear time-varying}
\newacronym{lmi}{LMI}{linear matrix inequality}
\newacronym{iss}{ISS}{input-to-state stability}
\newacronym{c+i}{C+I}{consensus + innovations}
\newacronym{lf}{LF}{Lyapunov function}
\newacronym{guas}{GUAS}{Global Uniform Asymptotic Stability}
\newacronym{ges}{GES}{Global Exponential Stability}
\newacronym{uco}{UCO}{Uniform Complete Observability}
\newacronym{pe}{PE}{Persistency of Excitation}
\newacronym{cpe}{cPE}{cooperative Persistency of Excitation}
\newacronym{sdp}{SDP}{semi-definite program}
\newacronym{mae}{MAE}{mean absolute error}
\newacronym{hv}{HV}{high-voltage}
\newacronym{mv}{MV}{medium-voltage}
\newacronym{lv}{LV}{low-voltage}
\newacronym{coi}{COI}{center of inertia}
\newacronym{pfc}{PFC}{primary-frequency-controlled}
\newacronym{entsoe}{ENTSO\Hyphdash E}{European Network of Transmission System Operators for Electricity}
\newacronym{scada}{SCADA}{Supervisory Control and Data Acquisition}

\DeclareMathOperator{\R}{\mathbb{R}}

\DeclareMathOperator{\Rp}{\mathbb{R}_{>0}}
\DeclareMathOperator{\Rps}{\mathbb{R}_{\ge0}}

\DeclareMathOperator{\Co}{\mathbb{C}}

\DeclareMathOperator{\diag}{diag}

\begin{document}
	
	\title{Consensus + Innovations Approach for Online Distributed Multi-Area Inertia Estimation
		%\thanks{}
  }

	\author{Nicolai Lorenz-Meyer, Hans Würfel, and Johannes Schiffer% <-this % stops a space
	\thanks{N. Lorenz-Meyer is with the Brandenburg University of Technology Cottbus-Senftenberg, Cottbus, Germany (e-mail: lorenz-meyer@b-tu.de).}
	\thanks{Hans Würfel is with the Potsdam Institute for Climate Impact Research, Potsdam, Germany (e-mail: wuerfel@pik-potsdam.de).}
	\thanks{J. Schiffer is with the Brandenburg University of Technology Cottbus-Senftenberg, Cottbus, Germany and the Fraunhofer Research Institution for Energy Infrastructures and Geothermal Systems (IEG), Cottbus, Germany (e-mail: schiffer@b-tu.de).}
}

	\maketitle
	
	\begin{abstract}
The reduction of overall system inertia in modern power systems due to the increasing deployment of distributed energy resources is generally recognized as a major issue for system stability.
Consequently, real-time monitoring of system inertia is critical to ensure a reliable and cost-effective system operation.
Large-scale power systems are typically managed by multiple transmission system operators, making it difficult to have a central entity with access to global measurement data, which is usually required for estimating the overall system inertia.
  We address this problem by proposing a fully distributed inertia estimation algorithm with rigorous analytical convergence guarantees.  This method requires only peer-to-peer sharing of local parameter estimates between neighboring control areas, eliminating the need for a centralized collection of real-time
measurements. We robustify the algorithm in the presence of typical power system disturbances and demonstrate its performance in simulations based on the well-known New England IEEE-39 bus system. 
	\end{abstract}
	
	\section{Introduction}
	\label{sec:intro_dist_H_est}
		The worldwide energy transition is leading to a progressive replacement of conventional \glspl{sg} by power-electronic interfaced \glspl{der} \cite{dorfler_control_2023, paolone_fundamentals_2020, winter_pushing_2015}. %. \nlm{Remove? Many of these \glspl{der}, such as solar and wind, are inherently intermittent .}
		 \glspl{sg}, which physically provide instantaneous energy reserve through the inertia of their rotating mass, play a key role in smoothing out power fluctuations in the grid. While it is possible to control \glspl{der} to support the grid and contribute to the total inertia~\cite{dorfler_control_2023}, the vast majority of \glspl{der} installed nowadays do not provide inertia. Hence, the total inertia available in power systems decreases~\cite{heylen_challenges_2021}. At the same time, new demand-response technologies and increasingly complex loads are installed, resulting in higher and reverse power flows~\cite{winter_pushing_2015}. Consequently, power systems are more frequently operated closer to the stability limits, and the overall dynamics become faster \cite{milano_foundations_2018}.
 The \gls{entsoe} recently identified the decrease in system inertia as the most critical power system stability issue for the continental European grid \cite{christensen_et_al_high_2020}. As such, tracking the total inertia and assessing the stability of the system becomes increasingly challenging for \glspl{tso} in the uncertain and deregulated environment of modern power systems \cite{ulbig_impact_2014, orum_future_2015}.
	This problem is further aggravated by the fact that large multi-area power systems are typically operated by a number of independent \glspl{tso}, each responsible for a separate control area. Since estimating the total inertia usually requires measurement data from the entire system, cooperation among \glspl{tso} becomes imperative.
	
The development of methods for estimating the power system inertia has recently received considerable research attention. The presented schemes follow a centralized approach and can be broadly categorized into offline, online, and forecasting methods~\cite{heylen_challenges_2021}. For in-depth discussions, the reader is referred to~\cite{heylen_challenges_2021,prabhakar_inertia_2022, tan_power_2022} for recent reviews on the subject. While many promising solutions have been proposed---to the best of the author's knowledge---no online distributed method with rigorous analytical convergence guarantees and robustness in the presence of disturbances has been presented. In such a way, the application in a practical setting and during different grid operation conditions can be facilitated as only peer-to-peer sharing of local parameter estimates between neighboring control areas is needed, avoiding a central entity and sharing of real-time measurements. This significantly reduces the communication infrastructure required and avoids the potential disclosure of sensitive data.%, such as transmission grid details. 

In this context, we propose an online robust \gls{c+i}-based distributed parameter estimator (see \cite{lorenz-meyer_robust_2024} for a theoretical analysis of the algorithm) for the inertia in multi-area power systems. More precisely, our contributions are three-fold:
	\begin{itemize}%[noitemsep,topsep=0pt]
	\item 	Based on the \gls{coi} dynamics of each control area, we derive \glspl{linre} depending only on local measurements, which enable the subsequent distributed parameter estimation scheme.
	
	\item We propose a method to estimate in real-time the inertia constants of all control areas and the total inertia of the multi-area power system in a fully distributed manner. 
  Moreover, we analytically establish the global convergence of the parameter estimates to the true parameter vector under standard assumptions. 
 Lastly, we robustify the algorithm in the presence of typical disturbances in power systems, such as parameter variations, measurement noise, and disturbances in the communication channels. 
	
	\item A simulation study based on the well-known New England IEEE-39 bus system \cite{hiskens_ieee_2013}, illustrates the effectiveness of the proposed algorithm. It is shown that the method accurately tracks the total inertia and the inertia constants of all control areas at each control area during regular grid operation, even in the presence of various disturbances.

\end{itemize}
	The remainder of this paper is organized as follows. In Section~\ref{sec:H_power_sys_model}, the mathematical model of a multi-area power system is presented. Based on that, a robust \gls{c+i}-based multi-area inertia estimator is introduced in Section~\ref{sec:C_I_H_estimator}. In Section~\ref{sec:H_est_simulation}, simulation results are presented. Lastly, in Section~\ref{sec:conclusions}, conclusions and a brief outlook on future work are provided.

	\section{Modeling of multi-area power systems}
	\label{sec:H_power_sys_model}
	
	%As in Chapter~\ref{ch:state_and_para_est}, 
	We consider a large-scale power system with a total of $\nsw>1$ interconnected control areas operated by different \glspl{tso} and comprised of $\Nsg$~$>$~$1$ \glspl{sg}. % Consequently, the overall power system has $\Nsw = \sum_{j=1}^{n} \Nsw_j$ \glspl{sg}, which are numbered from $1$ to $\Nsw$ \nlm{irgendwie ein komischer satz}. 
	The dynamic behavior of the $i$th \gls{sg} is represented by a standard swing equation model describing the electromechanical dynamics and given by (see, e.g., \cite{andersson_dynamics_2012}) 
	\begin{equation}
		\label{eq:modelSG_for_H}
		\begin{split}
			\ddeltasw_i(t) &= \Domegasw(t) = \omegasw_i(t)-\omega_\mrm{s}  ,\\
			\domegasw_i(t)&=\frac{\omega_\mrm{s}}{2\Hsw_i}\left(\Pmswi(t)-\Peswi(t)\right)  ,
		\end{split}
	\end{equation}
	where the subscript $i$ denotes the variables associated to the $i$th SG, with $\deltasw_i(t)\in \R$ denoting the rotor angle, $\omegasw_i(t) \in \Rps$ the shaft speed, $\Peswi(t)\in \Rps$ the electrical air-gap power, the mechanical power $\Pmswi(t)\in \Rps$, $\omega_\mrm{s} \in \Rps$ the nominal synchronous speed, and $\Hsw_i\in \Rp$ the inertia constant. To simplify the presentation, in the remainder of the paper, all powers and inertia constants are given in a common global per-unit system. It is assumed that the shaft speed $\omegasw_i$ is close to the nominal synchronous speed $\omega_s$, which is a usual assumption in power systems (see, e.g., \cite[Chapter 2]{andersson_dynamics_2012} and \cite[Chapter 5]{machowski_power_2008}). Moreover, assuming the stator resistance $\Rsg$ is zero, the electrical air-gap power $\Peswi(t)$ of each \gls{sg} can be approximated by the terminal electrical power $\Ptswi(t) \in \Rps$, which can be measured by the \gls{tso} (see, e.g., \cite{ghahremani_local_2016}). 
	
	%\begin{remark}
	%	Differing from Chapter~\ref{ch:decentralized_ad_obs}, no assumption is made on the electrical model of the \gls{sg}. Hence%, and to present the general form of the swing equation
	%	, it is assumed that the effects of the damper windings are modeled via the electrical model of the machine and, thus, influence the dynamics of the electrical air-gap power $\Peswi(t)$. Consequently, no additional term depending on the damping coefficient and the deviation from synchronous speed is considered in the model \eqref{eq:modelSG_for_H}. The damping coefficient is usually introduced in the \gls{sg} model to account for neglected asynchronous torques produced by the damper windings. As the damper windings are neglected when the \gls{sg} model is reduced to a fourth-order or lower-order model, an additional torque depending on the damping coefficient and the deviation from synchronous speed is introduced in these models \cite[Chapter 11]{machowski_power_2008}. %As no assumption on the electrical model of the \gls{sg} is made here, the form of the swing equation given in \eqref{eq:modelSG_for_H} is considered in this chapter. 
	%	If a third or fourth-order \gls{sg} model is used, the influence of the damping torque can be included in \eqref{eq:modelSG_for_H} as an additional term subtracted from the electrical air-gap power.
	%\end{remark}

	In the following, we consider that the \glspl{sg} of the power system are strongly coupled to each other. Thus, the \gls{coi} frequency can be used to represent the principal frequency dynamics of the overall power system. The \gls{coi} frequency $\omegacoi(t) \in \Rps$ is defined as (see, e.g., \cite[Chapter 2]{andersson_dynamics_2012})%, \cite{schiffer_online_2019}, and \cite[Chapter 2]{orum_future_2015}) 
	\begin{equation*}
		\omegacoi(t) = \frac{\sum_{i=1}^{\Nsg} \Hsw_i \omegasw_i(t)}{\sum_{i=1}^{\Nsg} \Hsw_i}.
	\end{equation*}
	As all inertia constants $\Hsw_i$ are given in the same global per-unit system, the total inertia of the power system $\Htot \in \Rp$ is calculated as
	\begin{equation*}
		%	\label{eq:Htot}
		\Htot = \sum_{i=1}^{\Nsg} \Hsw_i.
		%	\begin{split} H_\mrm{tot}^\mrm{S_B} &= \frac{\sum_{i} H_i^\mrm{S_B} S_{\mrm{B},i}}{\sum_{i} S_{\mrm{B},i}} = \frac{\sum_{i} H_i^\mrm{S_B} S_{\mrm{B},i}}{S_
				%		\mrm{B,tot}},
			%	\end{split}			
	\end{equation*}		
	%Hence, by introducing the relative \gls{coi} frequency $\xcoi \in \R$ as 
	%\begin{equation*}
	%	\xcoi(t) = \omegacoi(t)-\omega_\mrm{s},
	%\end{equation*}
	Hence, the dynamics of the \gls{coi} frequency can be expanded as
	\begin{equation}
		\label{eq_xcoi_central}
		\dxcoi(t) = \frac{\sum_{i=1}^{\Nsg} \Hsw_i \domegasw_i(t)}{\sum_{i=1}^{\Nsg} \Hsw_i} = \frac{\omega_\mrm{s}}{{2\Htot}}\left(\Pmtot(t)-\Petot(t)\right) ,
	\end{equation}
	where the total generated electrical power $\Petot(t)\in \Rps$ and the total mechanical power $\Pmtot(t)\in \Rps$ are defined as 
	\begin{equation*}
		\Petot(t) = \sum_{i=1}^{\Nsg} \Peswi(t) \quad \text{and} \quad \Pmtot(t) = \sum_{i=1}^{\Nsg} \Pmswi(t) ,
	\end{equation*}
	respectively.
	
	As large-scale power systems are usually operated by several independent \glspl{tso}, typically no central entity has access to real-time power and frequency measurements of the overall power system and, hence, estimating the total inertia constant $\Htot$ from \eqref{eq_xcoi_central} is difficult in a practical setting.
	Nevertheless, the same approach as detailed above can be followed for each control area within a multi-area power system. 
	Here, the \gls{coi} frequency can be used to represent the principal frequency dynamics of each control area. The \gls{coi} frequency of the $j$th area $\omegacoi_j(t) \in \Rps$ is given by
	\begin{equation}
		\label{eq:coi_j_area}
		\omegacoi_j(t) = \frac{\sum_{i \in \Nsw_j} \Hsw_i \omegasw_i(t)}{\sum_{i \in\Nsw_j} \Hsw_i},
	\end{equation}
	where the set of \glspl{sg} in the $j$th control area is denoted by $\Nsw_j$. 
	The total inertia of the $j$th area $\Hca_j \in \Rp$ follows from
	\begin{equation*}
		%	\label{eq:Htot}
		\Hca_j = \sum_{i \in \Nsw_j} \Hsw_i.
		%	\begin{split} H_\mrm{tot}^\mrm{S_B} &= \frac{\sum_{i} H_i^\mrm{S_B} S_{\mrm{B},i}}{\sum_{i} S_{\mrm{B},i}} = \frac{\sum_{i} H_i^\mrm{S_B} S_{\mrm{B},i}}{S_
				%		\mrm{B,tot}},
			%	\end{split}			
	\end{equation*}		
	%Hence, by introducing the relative \gls{coi} frequency of the $j$th area $\xcoi_j (t) \in \R$ as 
	%\begin{equation*}
	%	\xcoi_j(t) = \omegacoi_j(t)-\omega_\mrm{s},
	%\end{equation*}
	Consequently, the dynamics of the \gls{coi} frequency of the $j$th area can be expressed as
	\begin{equation}
		\label{eq:coi_frequency_j_area}
		\dxcoi_j(t) = \frac{\sum_{i \in \Nsw_j} \Hsw_i \domegasw_i(t)}{\sum_{i \in\Nsw_j} \Hsw_i} = \frac{\omega_\mrm{s}}{{2\Hca_j}}\left(\Pmtotj(t)-\Petotj(t)\right) ,
	\end{equation}
	where the total generated electrical power of the $j$th area $\Petotj(t)\in \Rps$ and the total  mechanical power of the $j$th area $\Pmtotj(t)\in \Rps$ are defined as 
	\begin{equation*}
		\Petotj(t) = \sum_{i \in\Nsw_j} \Peswi(t) \quad \text{and} \quad \Pmtotj(t) = \sum_{i \in\Nsw_j} \Pmswi(t) , 
	\end{equation*}
	respectively.
	
	If the total inertia constants in all areas are known, the total inertia of the overall power system can be calculated from  
	\begin{equation}
		\label{eq:H_tot}
		\Htot =\sum_{j = 1}^{\nsw}\Hca_j.
	\end{equation}		
	As the \gls{coi} frequency of the $j$th area \eqref{eq:coi_j_area} depends on the unknown inertia constants $\Hsw_i$, we approximate it by the average frequency of that control area $\xav_j(t) \in  \Rps$ (cf. \cite{schiffer_online_2019}) as 
	\begin{equation}
		\xav_j(t) = \frac{\sum_{i \in \Nsw_j}\omegasw_i(t) }{|\Nsw_j|},
	\end{equation}
	where $|\Nsw_j|$ denotes the cardinality of the set $\Nsw_j$, i.e., the number of \glspl{sg} within the $j$th control area.
	Then, the dynamics of the \gls{coi} frequency of the $j$th area \eqref{eq:coi_frequency_j_area} can be approximated by the dynamics of the average frequency as 
	\begin{equation}
		\label{eq:av_frequency_H_est}
		%	\dxav_j(t) = \aca \left(\Pmtotj(t)-\Petotj(t)\right) ,
		\dxav_j(t) = \frac{\omega_\mrm{s}}{{2\Hca_j}}\left(\Pmtotj(t)-\Petotj(t)\right) .
	\end{equation}
	%where $\aca =  \frac{\omega_\mrm{s}}{{2\Hca_j}}$.
	As detailed above, the electrical air-gap power $\Peswi(t)$ of each \gls{sg} can be approximated by the terminal electrical power $\Ptswi(t)$ and, hence, the total generated electrical power of the $j$th control area $\Petotj(t)\in \Rps$ can be approximated as the sum of the terminal electrical powers $\Ptswi (t)$ via %\nlm{(cf. \cite{ghahremani_local_2016})}
	\begin{equation*}
		\Petotj(t) = \sum_{i \in\Nsw_j} \Ptswi(t) .
	\end{equation*}
\begin{comment} The total electrical air-gap power of the $j$th control area $\Petotj(t)$ can be expressed as
	\begin{equation*}
		\Petotj(t) = \Peloadj(t) - \Peresj (t)+ \Pelossj(t) + \Peiaj(t) ,
	\end{equation*}
	where $\Peloadj \in \Rps$ denotes the total load demand of the $j$th control area, $\Peresj \in \Rps$ the total electrical power generated by \glspl{der} with grid-following control schemes in the $j$th control area, $ \Pelossj\in\Rps$ the total losses within the $j$th control area, and $\Peiaj \in \R$ the total interarea power flow from the $j$th control area to neighboring control areas. Note that the interarea power flow can be negative or positive depending on the direction of the power flow, i.e., leading into our out of the $j$th control area.
	\end{comment}
 
	In the following, it is assumed that the \glspl{tso} have access to local measurements from their control area. More precisely, the assumption on available measurements is summarized as follows.  
	\begin{assumption}
		\label{ass:measurements_H_est}
		The system operator of the $j$th control area has access to the signals $\xav_j(t)$, $\Pmtotj(t)$, and $\Petotj(t)$.
	\end{assumption}
 \begin{remark}
	The total mechanical power $\Pmtotj(t)$ of the $j$th control area can be composed of a constant and a time-varying part. The constant part depends on the (piecewise) constant power setpoints for each \gls{sg} within one control area, which are typically known to the respective \gls{tso}. The optional time-varying part is due to possible primary-frequency control of some of the \glspl{sg} and can, e.g., be modeled via a governor and turbine model \cite{schiffer_online_2019}.
	\end{remark}

	%\begin{remark}
	%	In \cite{schiffer_online_2019}, a centralized scheme is used to estimate the total inertia constant. It is assumed that a central entity has access to measurements of the overall power system. This assumption is relaxed in the approach presented in this chapter by assuming that each \gls{tso} can only access local measurements from its respective control area. 
	%	As done in \cite{schiffer_online_2019}, it is possible to further relax Assumption~\ref{ass:measurements_H_est} by dividing the \glspl{sg} into \gls{pfc} and uncontrolled generators and approximating \eqref{eq:av_frequency_H_est} by only considering the \gls{pfc} generators. This has the drawback of naturally introducing an offset in the estimated inertia constants due to the approximation. This relaxation is not included in this chapter but can be added swiftly following the approach presented in \cite{schiffer_online_2019}. 
	%\end{remark}

	\begin{remark}
		The frequency dynamics of today's transmission systems, e.g., the synchronous grid of continental Europe, is dominated by \glspl{sg}. Nevertheless, in modern power systems, \glspl{sg} are continuously phased out, and at the same time, power-electronics interfaced \glspl{der} based on renewable energy sources are implemented in large numbers. Thus, in the future, at least a part of the total inertia has to be provided by \glspl{der} through grid-forming control \cite{dorfler_control_2023}. This can be achieved, e.g., by controlling the \glspl{der} to follow the dynamics of the swing equation, as virtual synchronous machines \cite{darco_virtual_2013} or using droop control \cite{barklund_energy_2008, schiffer_conditions_2014} among others. Consequently, the frequency dynamics of \glspl{der} can be included in the \gls{coi} frequency dynamics \eqref{eq_xcoi_central} and \eqref{eq:coi_frequency_j_area}. 
		%	As the swing equation \eqref{eq:modelSG_for_H} only considers the mechanical dynamics of the \gls{sg}, the effects of the damper windings are modeled via the electrical model of the machine and, thus, influence the dynamics of the electrical air-gap power $\Peswi(t)$. 
		Differing from the \gls{sg} model \eqref{eq:modelSG_for_H}, the damping term in \glspl{der} under grid-forming control is due to the structure of the control scheme itself (see, e.g., \cite{scholl_design-_2021,schiffer_synchronization_2013}). Hence, it must be incorporated in the \gls{coi} frequency dynamics as an additional unknown constant, which can be included in the estimation method, e.g., following a similar approach as presented in \cite{lorenz-meyer_pmu-based_2020-1}.% Furhtermore, in such case the active power injection due to the primary frequency control of the gernating units $P_{\mrm{PFC},i}(t)$ could be modeled in dependence of the frequency deviation as a weighted sum of the turbine dynamics of the SGs and the $P(f)$ characteristic of the DERs.
	\end{remark}

	\section{An online robust C+I-based distributed multi-area inertia estimator}
	\label{sec:C_I_H_estimator}
	In the following, a distributed estimation scheme is developed to reconstruct in real-time the inertia constants $\Hca_j$ of all control areas as well as the total inertia constant of the interconnected power system $\Htot$ at each control area using the measurements introduced in Assumption~\ref{ass:measurements_H_est}. 
	
			%To facilitate the development of a distributed parameter estimator in the following,
			For this, communication between the control areas is required and it is assumed that neighboring control areas communicate their local parameter estimates with each other in a peer-to-peer fashion. The communication structure among the $\nsw$ control areas is represented by an undirected, time-dependent graph $\Ghcal(t) = \left(\Vhcal, \Ehcal(t)\right)$, where $\Vhcal= \{1,2,...,\nsw\}$ is the set of nodes, $\Ehcal(t)\subseteq\Vhcal \times \Vhcal$ is the set of edges at time $t$, and the cardinality of $\Ehcal(t)$ is denoted by $\nhe(t)$. In this way, communication failures can be included in the model as the communication structure is allowed to change over time. The oriented incidence matrix $\Dh(t) \in \mathbb{R}^{\nsw \times \nhe(t) }$ is constructed by assigning an arbitrary orientation to the edges of the graph $\Ghcal(t)$ at time $t$, and is defined element-wise as $\Dh_{ij}(t) = 1$, if $i$ is the source of the $l$th edge, $\Dh_{ij}(t) = - 1$, if $i$ is the sink of the $l$th edge, and $\Dh_{ij}(t) = 0$ otherwise.	
	The Laplacian matrix $\Lh(t) \in  \mathbb{R}^{\nsw\times \nsw}$ of the graph $\Ghcal(t)$ at time $t$ is defined as \cite{bullo_lectures_2020}
	\begin{equation}
		\label{eq:L_H}
		\Lh(t) = \Dh(t)\Dh^\top(t).
	\end{equation} 
	
To obtain \glspl{linre}, the unknown constants $\aca_j \in \Rp$ of the $j$th control area and $\atot \in \Rp$ of the overall power system are introduced as\footnote{For the remainder of this work, the signals' time arguments are omitted whenever the time dependency is clear from the context.}
	\begin{equation}
		\label{eq:a_H_Est}
		\aca_j =  \frac{\omega_\mrm{s}}{{2\Hca_j}}, \quad \atot =  \frac{\omega_\mrm{s}}{{2\Htot}}=  \frac{1}{{\sum_{j = 1}^{\nsw}\aca_j}},
	\end{equation}
	respectively.
	Hence, by defining the linear second-order filter $\Fav \in \Co$ as
	\begin{equation*}
		\Fav (s)= \frac{\lambdaav1 \lambdaav2}{(\lambdaav1+s)(\lambdaav2+s)},
	\end{equation*}
	with $s $ being the complex frequency domain parameter and $\lambdaavi \in \Rp, \ i=\{1,2\}$ denoting tuning parameters, a local \gls{linre} can be formulated for each control area by applying $\Fav$ to \eqref{eq:av_frequency_H_est}, i.e., 
	\begin{equation} \label{eq:LRE_H}
		\Fav\left[s\left[\xav_j\right]\right] =
		\Fav\left[\Pmtotj-\Petotj\right] \aca_j.
	\end{equation}
	As the aim is to estimate all inertia constants of all control areas, the parameter vector $\thetah \in \R^{\nsw}$ to be estimated is defined as follows
	$$\thetah = \begin{bmatrix}\aca_1 & \aca_2 &\dots & \aca_{\nsw}
	\end{bmatrix}^\top .$$
	Furthermore, we define the local output $\yh_j \in \R$ of the $j$th control area as 
	$y_j \coloneqq 	\Fav\left[s\left[\xav_j\right]\right],  $
	and $\nuh_j \in \R$ as
	$\nuh_j$~ $\coloneqq$~$ 		\Fav\left[\Pmtotj-\Petotj\right] .  $
	Then, the local \gls{linre} \eqref{eq:LRE_H} for the $j$th control area can be expressed in dependence of the global parameter vector $\thetah$ in the form of the following \gls{linre} 
	\begin{equation*}
		\yh_j =
		\Ch_j  \thetah,
	\end{equation*}
	where $\Ch_j \in \R^{1\times \nsw}$ denotes the regressor of the $j$th control area and is defined as  %Similarly, the \glspl{linre} for the $ \nsw-1$ remaining control areas can be obtained.  
	\begin{equation*}
		\Ch_j \coloneqq \begin{cases}
			\begin{bmatrix}
				\nuh_1 & {0}^\top_{(\nsw-1)}	\end{bmatrix} & \mrm{for} \
			j = 1, \vspace{0.1cm}
			\\ 
			\begin{bmatrix}
				{0}^\top_{(\nsw-1)}&	\nuh_{\nsw} \end{bmatrix} & \mrm{for}  \ j= \nsw, \vspace{0.1cm} \\ 
			\begin{bmatrix}
				{0}^\top_{(j-1)}&	\nuh_j & {0}^\top_{(\nsw-j)}	\end{bmatrix} & \mrm{otherwise},
		\end{cases} 
	\end{equation*}
	with ${0}_h$ representing a $h\times1$ vector of zeros.
	
	The goal is to estimate consistent parameters $\thetah$ at all control areas, which can be achieved by a \gls{c+i}-type distributed parameter estimation algorithm \cite{chen_distributed_2014, lorenz-meyer_robust_2024}. 
	With this algorithm, each control area $j$ continuously updates its local estimate $\hthetah_j \in \R^{\nsw}$ using 
	\begin{equation}
		\label{eq:C+I_alg_H}
		\dhthetah_j = -\alphah\Gammah_j \sum_{k\in\Nhcal_j}(\hthetah_j-\hthetah_k) - \Gammah_j \Ch_j^\top\left(\Ch_j \hthetah_j-\yh_j\right),
	\end{equation}
	where $\Nhcal_j$ denotes the set of neighbors of the $j$th control area at time $t$, $\Gammah_j \in \mathbb{R}^{\nsw\times \nsw}$ is the gradient descent gain matrix of the $j$th control area, and $\alphah \in \mathbb{R}$ is the additional consensus gain. Notably, this algorithm only requires the communication of the local inertia estimates to neighboring control areas without the need to disclose the underlying local frequency and power measurements.
	
		For establishing the convergence properties of the parameter estimates $\hthetah_j  \  \forall j=\{1,\cdots,\nsw\}$ in \eqref{eq:C+I_alg_H}, the subsequent standard assumptions are imposed.   
	\begin{assumption} \
		\label{ass:cPEH}
		\begin{enumerate}[noitemsep,topsep=0pt]
			\item The cooperative Persistency of Excitation condition is fulfilled, i.e., there exist positive constants $\Timeh$, $\cPEuh \ge\cPElh > 0$, all independent of $t$, such that for all $t\ge \Timeh$ it holds that
			\begin{equation*}
				\cPEuh I_{N} \ge \int_{t-\Timeh}^{t}\sum_{j=1}^{\nsw}\Ch_j^\top(s){\Ch_j}(s)\mathrm{d}s \ge \cPElh I_{N}  .
			\end{equation*}
			\item All the local regressors are uniformly upper-bounded in the norm, with the upper bound given by $\rh_2>0$, i.e., $
			\|\Ch_j^\top(t) \Ch_j(t)\| \le  \rh_2 \quad \forall\,t\geq 0 \quad \forall j=\{1,\cdots,\nsw\}.
			$
			\item The graph $\Ghcal(t)$ is connected on average, meaning that $\int_{t-\Timeh}^{t}\Lh(s)\mathrm{d}s$, with $\Timeh$ as in Assumption~\ref{ass:cPEH}.1, has only one zero eigenvalue $\lambdah1$, and all remaining ones are positive and accept a lower bound denoted by the constant $0<\lambdahunder\le \lambdahk(t)  \quad  \forall k \ge 2$. Furthermore, the time-varying Laplacian $\Lh(t)$ in \eqref{eq:L_H} is uniformly upper-bounded in the norm, with upper bound $\rh_3>0$, i.e., $\|\Lh(t)\| \le \rh_3 \quad \forall\,t\geq 0.	
			$
			\item 	For each control area, its gradient descent gain matrix $\Gammah_j$ in \eqref{eq:C+I_alg_H} is symmetric and positive definite, with the upper bound given by $\rh_1 >0$, i.e., $
			\|\Gammah_j\| \le  \rh_1 \quad \forall j=\{1,\cdots,\nsw\}.
			$
			Moreover, the additional consensus gain $\alphah$ in \eqref{eq:C+I_alg_H} is chosen to be strictly positive.
		\end{enumerate} 
	\end{assumption}
 In the literature, Assumption~\ref{ass:cPEH} is widely used and is in line with standard practice (see, e.g., \cite{rueda-escobedo_strong_2021} for Assumption~\ref{ass:cPEH}.2, \cite{chen_distributed_2014} for Assumption~\ref{ass:cPEH}.1 and Assumption~\ref{ass:cPEH}.3, while Assumption~\ref{ass:cPEH}.4 can always be satisfied by proper design).
	With Assumption~\ref{ass:cPEH}, global convergence of the estimates $\hthetah_j \ \forall j=\{1,\cdots,\nsw\}$ in \eqref{eq:C+I_alg_H} to the true parameter vector $\theta$, follows by invoking \cite[Theorem 1]{lorenz-meyer_robust_2024}. 

	From \eqref{eq:a_H_Est}, the online estimate of the unknown parameter vector at the $j$th control area $\hthetah_j$ can be directly used to estimate the unknown constant of the overall power system $\atot_j $ at the $j$th control area via 
	\begin{equation*}
		\label{eq:Htot_hat}
		\hatot_j =  \frac{1}{\sum_{i = 1}^{\nsw}\frac{1}{\hthetah_{j,i}}}, 
	\end{equation*}
	where $\hthetah_{j,i} \in \R$ denotes the $i$th component of the parameter estimate at the $j$th control area.
%\marginnote{	As \eqref{eq:Htot_hat} describes an algebraic relation between $\hatot_j$ and $\hthetah_j$, and $\hthetah_j$ converges to the correct values as shown in Theorem \ref{th:ges_H}, also $\hatot_j$ converges to the correct value. }
	Hence, at each control area, the inertia constants $\Hca_j$ of all control areas as well as the total inertia constant of the interconnected power system $\Htot$ can be reconstructed accurately by employing \eqref{eq:a_H_Est}.
	
Moreover, following \cite[Corollary~4]{lorenz-meyer_robust_2024}, the gains of the distributed multi-area inertia estimator $\bGammah = \diag\left(\Gammah_1,\cdots, \Gammah_{\nsw}\right) $ and $\alphah$ in \eqref{eq:C+I_alg_H} can be tuned systematically. In this way, the $\mathcal{L}_2$-gain from a disturbance input to a specified performance output can be minimized utilizing the semi-definite program provided in \cite[Corollary~4]{lorenz-meyer_robust_2024}. Thus, the proposed algorithm \eqref{eq:C+I_alg_H} can be robustified in the presence of typical disturbances, such as measurement noise, disturbances in the communication channels, and parameter variations. Due to space limitations, the exact procedure is not explained here, but follows the one described in \cite{lorenz-meyer_robust_2024}. Hence, the interested reader is referred to \cite{lorenz-meyer_robust_2024} for further details.

\section{Simulation results}
\label{sec:H_est_simulation}
\begin{figure}
	\centering
	\includegraphics[width=0.8\linewidth]{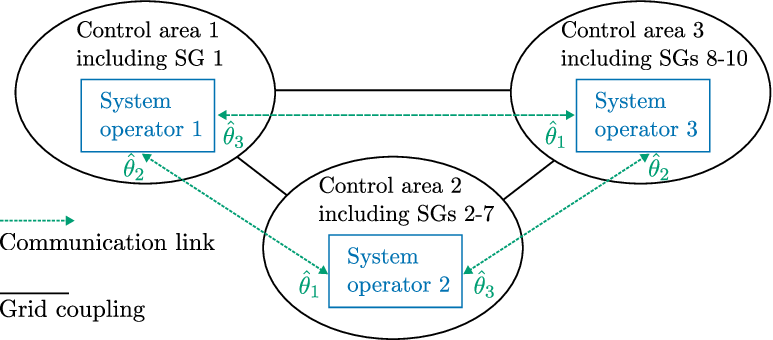}
	\caption{Schematic representation of the considered control areas and the communication topology. }
	\label{fig:H_est_communication}
\end{figure}

In this section, simulation results utilizing the \gls{c+i}-based multi-area inertia estimator introduced in \eqref{eq:C+I_alg_H} are presented. For this, the well-known New England IEEE 39 bus system \cite{hiskens_ieee_2013} is used with the \glspl{sg} modeled via a 9-dimensional model including automatic voltage regulators and power system stabilizers. The power system is divided into three control areas. The \Gls{sg} 1 models the interconnection to an external power grid and, therefore, has a very large inertia constant \cite{linaro_continuous_2023}. Hence, control area 1 is assumed to be the external grid and
contains only \gls{sg} 1. Control area 2 comprises SGs 2-7, and control area 3 contains SGs 8-10. A schematic representation of the considered control areas and the communication topology is depicted in Figure~\ref{fig:H_est_communication}.

To validate the performance of the method during regular grid conditions, load variations are introduced. %As in Section~\ref{sec:load_variations_SG}, 
The resulting frequencies at all buses are in the range of $60\pm0.05$~Hz, which is consistent with the typical operation of transmission grids \cite{weissbach_verbesserung_2009}.
%\begin{figure}[h!]
%\centering
%\includegraphics[width=1\linewidth]{./figures/IEEE39_bus_multi_area}
%\caption{Allocation of the SGs to the three control areas considered in the New England IEEE 39 bus system. }
%\label{fig:IEEE39Bus_control_areas}
%\end{figure}
The available measurements satisfy Assumption~\ref{ass:measurements_H_est}, i.e., the system operator of the $j$th control area can measure $\xav_j$, $\Pmtotj$, and $\Petotj$. Furthermore, initially, all areas communicate their local inertia estimates with all other areas. %Thus, the communication topology is modeled by a complete graph. %A schematic representation of the considered control areas and the communication topology is depicted in Figure~\ref{fig:H_est_communication}.
%\begin{figure}[tb]
%\centering
%\includegraphics[width=0.8\linewidth]{./figures/Dist_setup_H_est}
%\caption{Schematic representation of the considered control areas and the communication topology. }
%\label{fig:H_est_communication}
%\end{figure}

To validate the performance of the proposed inertia estimator, we consider a scenario in which the inertia constant at the control area 1 is slowly time-varying. As control area 1 models the interconnection to an external grid, it is assumed that not only \glspl{sg} but also modern \glspl{der} contribute to the total inertia of this sub-grid. Since modern \glspl{der} do not provide inertia through the physical inertia of the rotating mass as in the case of \glspl{sg}, but through the control scheme applied in the form of virtual inertia, their inertia contribution may not be constant or even designed to be time-varying \cite{misyris_robust_2018,chen_adaptive_2020,liu_-line_2021}. Additionally, the inertia constant of control area 1 is reduced stepwise at $t= 40$ s, resembling, e.g., a disconnection of a large \gls{sg}. Furthermore, the communication is disturbed by removing the communication link between control area 2 and control area 3 for $t\ge 5$ s. Nevertheless, the graph modeling the communication topology remains connected. Lastly, we consider three cases with varying measurement noise: A nominal, i.e., noisefree case, a case with zero mean Gaussian noise, and a case with zero mean Laplacian noise added to the measurements. In \cite{wang_assessing_2018}, Laplacian noise was recommended to simulate realistic measurement errors. The signal-to-noise ratio of the measurement noise is set to $58$ dB and $95$ dB for the active power and frequency measurements, respectively. These values were determined to resemble the noise power in real-world applications by analyzing measurements from the German extra-high voltage grid obtained in \cite{lorenz-meyer_dynamic_2023}. To achieve robust performance in the presence of these disturbances, the gains of the algorithm are tuned using \cite[Corollary~4]{lorenz-meyer_robust_2024}. %The optimization problem is solved with the Julia package JuMP \cite{lubin_jump_2023}, the semi-definite program with MOSEK \cite{mosek_aps_mosek_2023}. 
The optimal gains found with this procedure are $\bGammahopt= 2.45I_{\nsw^2}$ and $\alphahopt = 0.4$.
All simulations are carried out using MATLAB and the Julia programming language. 

\begin{figure}
  \subfloat[Case 1: Noisefree measurements.] {%
	\includegraphics[width=\linewidth]{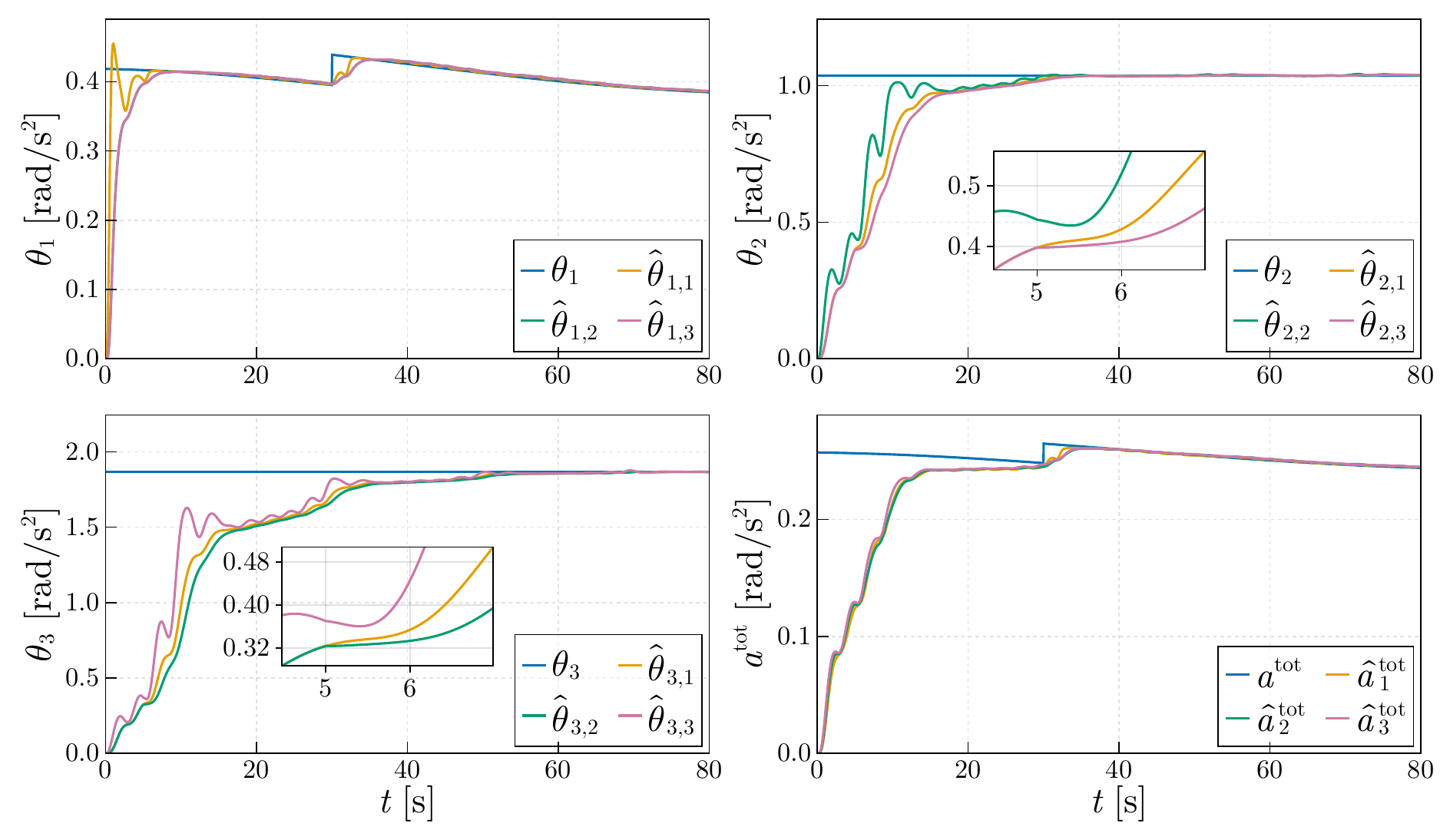}
	\label{fig:H_est_no_noise_S2}
  }
  \hfil
  \subfloat[Case 2: Measurements disturbed by zero mean Gaussian noise.] {%
	\includegraphics[width=\linewidth]{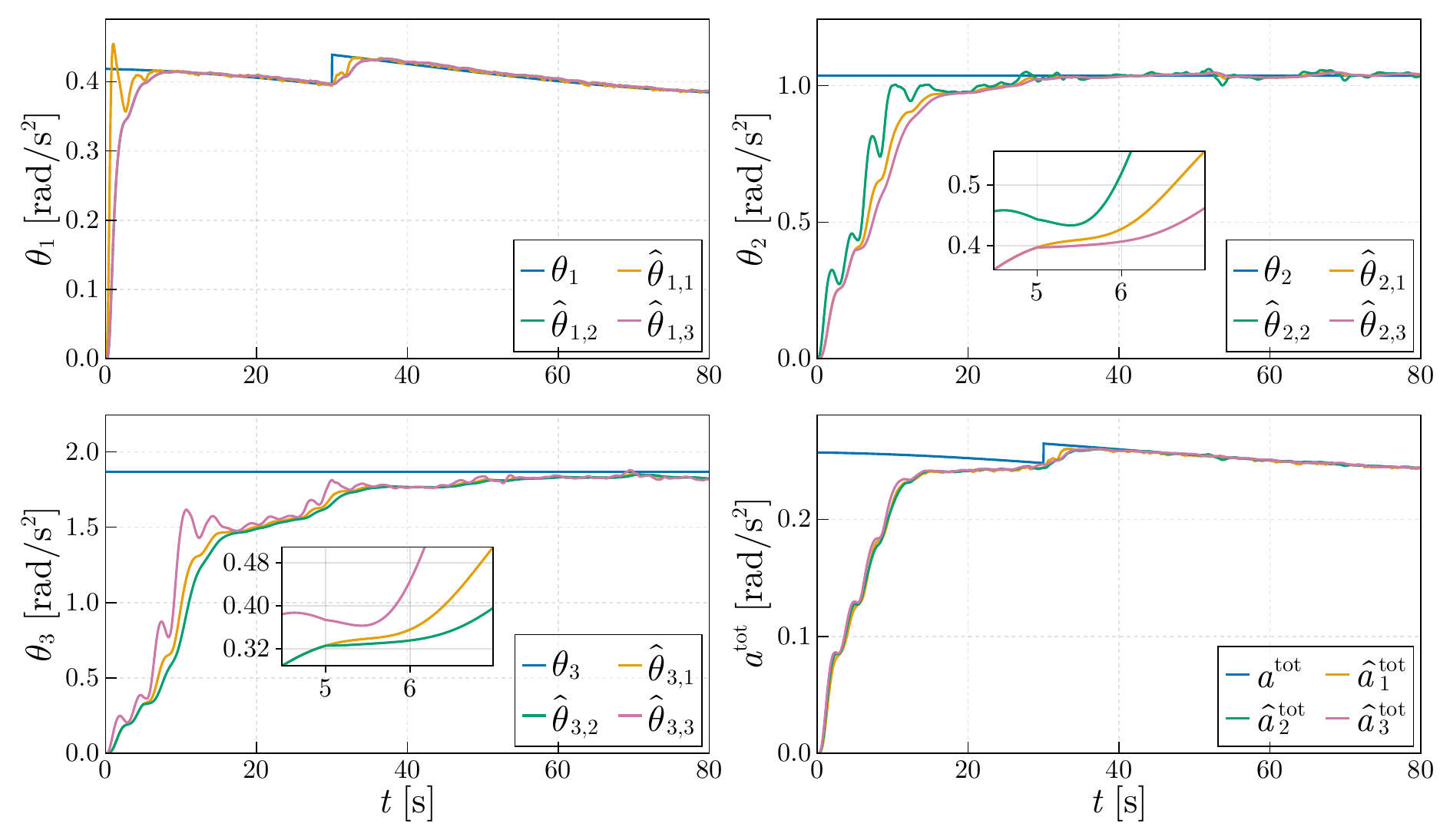}
	\label{fig:H_est_gaussian_S2}
  }
  \hfil
  \subfloat[Case 3: Measurements disturbed by zero mean Laplacian noise.] {%
	\includegraphics[width=\linewidth]{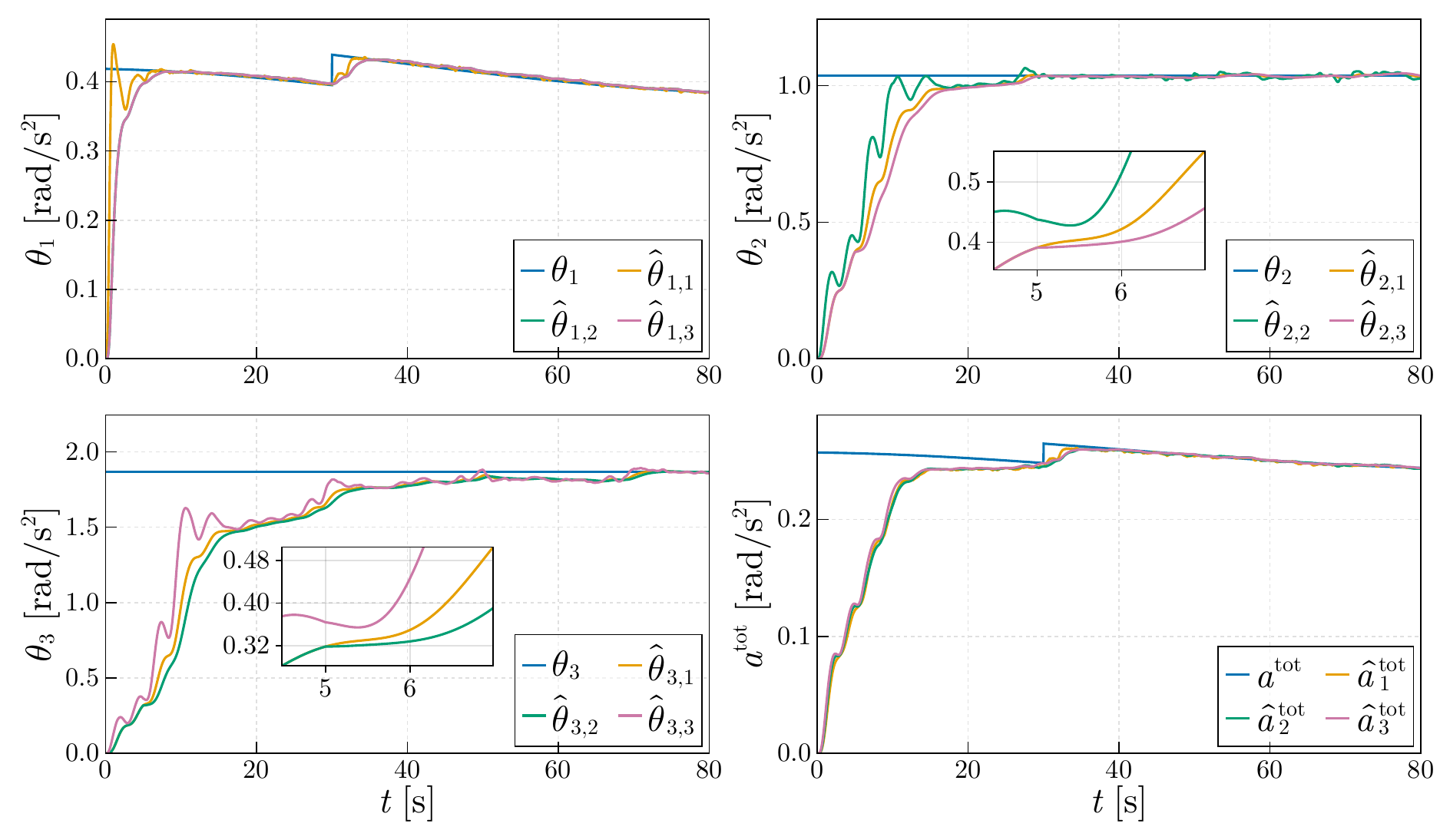}
	\label{fig:H_est_load_laplace_S2}
  }
	% \begin{subfigure}{\linewidth}
	% 	\centering
	% 	\includegraphics[width=\linewidth]{./figures/fig_estimated_parameters_CI_no_noise_S2.pdf}
	% 	\caption{Case 1: Noisefree measurements.}
	% 	% \vspace{0.2cm}
	% 	\label{fig:H_est_no_noise_S2}
	% \end{subfigure}

	% % \vspace{0.5cm}
	% \begin{subfigure}{\linewidth}
	% 	\centering
	% 	\includegraphics[width=\linewidth]{./figures/fig_estimated_parameters_CI_gaussian_noise_S2.pdf}
	% 	\caption{Case 2: Measurements disturbed by zero mean Gaussian noise.}
	% 	% \vspace{0.2cm}
	% 	\label{fig:H_est_gaussian_S2}
	% \end{subfigure}

	% \begin{subfigure}{\linewidth}
	% 	\centering
	% 	\includegraphics[width=\linewidth]{./figures/fig_estimated_parameters_CI_laplacian_noise_S2.pdf}
	% 	\caption{Case 3: Measurements disturbed by zero mean Laplacian noise.}
	% 	% \vspace{0.2cm}
	% 	\label{fig:H_est_load_laplace_S2}
	% \end{subfigure}

	\caption{Distributed inertia estimation in the presence of load variations with time-varying inertia at control area 1 and a loss of the communication link between control area 2 and 3 for $t\ge5$ s. }
	\label{fig:H_est_S2}
\end{figure}

The simulation results for all three cases are shown in Figure~\ref{fig:H_est_S2}. It can be seen that the distributed multi-area inertia estimator, in \eqref{eq:C+I_alg_H}, can accurately estimate the unknown parameters of all control areas and of the overall power system at all control areas. 
Initially and after the step change at $t=40$ s, the estimation of the time-varying parameter of control area 1 converges very quickly. The convergence of the estimation of the unknown parameters of control areas 2 and 3 takes slightly longer. The disturbance in the communication, i.e., the failure of the communication link between control areas 2 and 3 for $t \ge 5$ s, yields a delay in the estimation, which is most visible in the estimation of the unknown parameters of control areas 2 and 3. As seen from the zoomed-in plot at around $t=5$ s, the values of $\hthetah_{2,1}$ and $\hthetah_{2,2}$ as well as of $\hthetah_{3,1}$ and $\hthetah_{3,2}$ deviate from each other as the information can not be shared via the lost communication link anymore and travels a further way, i.e., via control area 1. As the graph describing the communication topology remains connected even after the loss of the communication link, the estimates still converge at all control areas. 
Overall, the robust distributed multi-area inertia estimator performs very well in all cases and is capable of correctly estimating the unknown parameters even in the presence of disturbances.

\section{Conclusions and future work}
\label{sec:conclusions}
In this work, we presented a novel \gls{c+i}-based online robust distributed inertia estimation method for multi-area power systems with rigorous analytical convergence guarantees. We are not aware of any other fully distributed scheme addressing this problem, i.e., without requiring a central entity with access to global measurement information. This is especially relevant as large-scale interconnected power systems are usually operated by multiple independent \glspl{tso}. Therefore, having a centralized inertia estimation entity may be difficult in a practical setting due to the potential disclosure of sensitive data and because it requires an elaborate infrastructure for real-time sharing of measurements. Instead, the proposed method requires only peer-to-peer sharing of local parameter estimates between neighboring control areas. In this way, sharing of the actual measurement data is not necessary.

More specifically, we derived a \gls{linre} from the \gls{coi} frequency model for each control area, which allowed us to apply a \gls{c+i}-based distributed parameter estimator to this problem setup. 
We analytically proved the \gls{ges} of the origin of the resulting error dynamics and robustified the algorithm in the presence of typical disturbances in power systems. Lastly, the algorithm was validated using simulation results based on the well-known New England IEEE 39 bus system. Here, the method could accurately reconstruct the inertia constants in the presence of communication disturbances, measurement noise, and variations in the inertia constant of the external grid. 

In the future, we plan to validate the method in a large-scale simulation study based on a dynamic 1013-machine \gls{entsoe} model through several test cases.

	\bibliographystyle{ieeetr}
\bibliography{Diss_bib}

\end{document}